\documentclass[a4paper,11pt]{article}
\usepackage{pos}
\usepackage{amsmath,amssymb,amstext}
\usepackage{subfigure}

\usepackage{float}

\usepackage{multirow}
\usepackage[utf8]{inputenc}
\usepackage{lmodern}

\usepackage{ etoolbox, xparse}

\usepackage{environ}

\usepackage{hyperref}

\usepackage{graphics,graphicx}
\usepackage{sidecap}
\usepackage[version=4]{mhchem}
\sidecaptionvpos{figure}{t}

\usepackage{wrapfig, framed}

\usepackage{xcolor}

\newcommand{\rj}{R_{\rm jet}}

\newcommand{\geff}{\Gamma_{\rm eff}}

\title{Ultra-High-Energy Cosmic Rays and Neutrinos from Relativistic Jets of Active Galactic Nuclei}
 \ShortTitle{UHECRs and Neutrinos from AGN Jets}

\author*[a,b,c]{Rostom Mbarek}
\author[a,c]{Damiano Caprioli}
\author[d,e,f,g]{Kohta Murase}

\affiliation[a]{University of Chicago, Department of Astronomy \& Astrophysics,\\
  5640 S Ellis Ave., Chicago, IL 60637, USA}

\affiliation[b]{Kavli Institute for Cosmological Physics, The University of Chicago, Chicago, IL 60637, USA}

\affiliation[c]{Enrico Fermi Institute, The University of Chicago, Chicago, IL 60637, USA}

\affiliation[d]{Department of Physics, The Pennsylvania State University, University Park, Pennsylvania 16802, USA}
\affiliation[e]{Department of Astronomy \& Astrophysics, The Pennsylvania State University, University Park, Pennsylvania 16802, USA}
\affiliation[f]{Center for Multimessenger Astrophysics, Institute for Gravitation and the Cosmos, The Pennsylvania State University, University Park,
Pennsylvania 16802, USA}
\affiliation[g]{Center for Gravitational Physics, Yukawa Institute for Theoretical Physics, Kyoto, Kyoto 606-8502 Japan}

\emailAdd{rmbarek@uchicago.edu}
\emailAdd{caprioli@uchicago.edu}
\emailAdd{murase@psu.edu}

\abstract{In \cite{mbarek+19}, we laid the groundwork for studying the \emph{espresso} paradigm \citep{caprioli15}, a reacceleration mechanism to boost galactic CRs to UHECR levels.
Our bottom-up approach uses realistic 3D MHD simulations of relativistic AGN jets and accounts for all of the crucial ingredients of a universal acceleration theory: injection, acceleration, and escape in  realistic environments. Our results are consistent with the main features of UHECR spectra, i.e., power-law slopes, chemical composition, and anisotropy. 
In \cite{mbarek+21}, we refine our model by including sub-grid particle scattering to model small-scale magnetic turbulence that cannot be resolved by MHD simulations, constraining for the first time one crucial but hard-to-model ingredient, and allowing us to establish the relative importance of \emph{espresso} and stochastic shear acceleration in relativistic jets. Our framework also enables us to analyze high-energy neutrinos produced from our accelerated UHECRs considering the effects of external photon fields, and to incorporate nucleus photodisintegration. The spectra we obtain are consistent with the picture drawn by observations with Auger, Telescope Array, and IceCube observatory.}

\FullConference{37$^{\rm{th}}$ International Cosmic Ray Conference (ICRC 2021)\\
		July 12th -- 23rd, 2021\\
		Online -- Berlin, Germany}

\begin{document}
\maketitle

\section{Introduction}

Ultra-High-Energy Cosmic rays (UHECRs) were first detected decades ago, and recently, the Pierre Auger Observatory has measured their spectrum in unprecedented detail \cite{auger17_coll}.
Yet, the production mechanism and astrophysical sources of UHECRs are still much debated. Acceleration models hardly go beyond back-of-the-envelope estimates of the maximum energy achievable, i.e. the Hillas limit \citep[e.g.][]{hillas84}, in astrophysical sources. Among such sources, the jets of active galactic nuclei (AGNs) are regarded as very promising candidates:  
they easily satisfy the Hillas limit up to the highest energies for iron nuclei and have luminosities that can sustain the UHECR energy injection rate.

We examine a promising theoretical framework based on the \emph{espresso} acceleration mechanism \cite{caprioli15,mbarek+19,mbarek+21}, which has the potential to satisfy the requirements of a comprehensive theory of UHECR acceleration in AGN jets. 
Such a framework accounts for: 
i) the specifics of particle injection into the source;
ii) a general acceleration mechanism, independent of poorly-constrained environmental paramters; 
and iii) the properties of the released spectra, chemical composition, and anisotropy. 
The theoretical basis of the \emph{espresso} mechanism was outlined in \cite{caprioli15} as a one-shot acceleration mechanism:
CR \emph{seeds} accelerated in supernova remnants up to PeV energies penetrate into AGN relativistic jets and generally receive a boost in energy of a factor of $\sim \Gamma^2$, where $\Gamma$ is the Lorentz factor of the relativistic flow.
Unlike stochastic acceleration mechanisms, \emph{espresso}-accelerated particles do not undergo multiple energy gain/loss cycles; 
more importantly, acceleration does not depend on highly uncertain parameters, such as the diffusion rate in the local magnetic turbulence.

In \cite{mbarek+19}, we used a bottom-up approach for investigating the \emph{espresso} framework while keeping parametrizations at a minimum.
More precisely, we studied particle injection and acceleration in high-resolution magnetohyrodynamic (MHD) simulations of relativistic jets run with {\tt PLUTO}\cite{mignone+12}. 
We found that particles are accelerated up to the jet's Hillas limit via one/two \emph{espresso} shots \emph{without} any added diffusion, further reiterating \emph{espresso}'s generality. 
In that sense, we concluded that \emph{espresso} is a generic acceleration mechanism that can explain the UHECR spectrum without resorting to particular assumptions. 

MHD jet simulations can self-consistently capture the large scale fluctuations in magnetic fields, but below the grid resolution, the potential role of smaller-scale turbulence is unaccounted for. 
On the other hand, stochastic acceleration processes may hinge on pitch-angle scattering produced by such unresolved scattering and be important in realistic AGN jets. 

Moreover,  the powerful radiation fields from the AGN broad-line region, dusty torus, or blazar zone in the environs of the jet could substantially alter the spectra of heavier UHECRs and produce lighter elements and neutrinos.
A UHECR theory should also include nucleus photodisintegration and possibly come with a prediction for the flux of high-energy neutrinos produced in the source. 

To address these matters, we add two main pieces to our self-consistent framework:
\begin{enumerate}
    \item We propagate particles on the grid with a standard Boris pusher \cite[e.g.,][]{birdsall+91}, while prescribing a finite probability per unit time for particles to change their pitch angle with respect to the local magnetic field. This determines a sub-grid scattering (SGS) rate that complements the one due to large-scale MHD fluctuations. Such a recipe enables us to probe the potential role of stochastic acceleration in AGN jets.
    \item We prescribe jet photon fields based on methods presented in \citep{murase+14} and use a Monte Carlo approach to quantify photodisintegration and its effects on the final UHECR spectrum. In addition, we calculate the expected fluxes of Ultra-High-Energy (UHE) neutrinos above $10^{17}$eV and of high-energy neutrinos, comparing results with  IceCube measurements.
\end{enumerate}

\section{Espresso and stochastic acceleration in AGN jets}\label{sgs}

\subsection{Sub-grid scattering prescription}
We propagate particles in a 3D relativistic MHD jet simulation performed with {\tt PLUTO} (see \cite{mbarek+19} for details about the setup), which includes adaptive mesh refinement \citep{mignone+10}. 
We augment our Boris-pusher-integrated trajectories with a Monte Carlo treatment of pitch-angle scattering, and span an array of scattering rates across two limiting cases: 
i) the limit in which diffusion occurs in the Bohm regime, such that the mean free path for pitch-angle scattering is as small as the particle's gyroradius, and ii) the limit of no SGS scattering, as in the simulations of \citep{mbarek+19}. 
We introduce a diffusion coefficient $D$ that depends on the particle rigidity and local magnetic field such that $D \left( \mathcal{R} \right) \equiv \frac{\kappa}{3} c \mathcal{R}\left( E, q,B \right)$, where $\kappa$ characterizes the number of gyroradii per sub-grid scattering and $\mathcal{R}$ is the particle gyroradius, a function of the local magnetic field $B \left( \textbf{r} \right)$ (see \citep{mbarek+21} for details). We compare a wide range of SGS rates by setting $\kappa = \infty $(no SGS), 1000, 100, 10, and 1 (Bohm regime) and investigate their effects on particle acceleration, energy gains, and anisotropy.

\begin{SCfigure}[][ht]
  \centering
  \caption{\protect\rule{0ex}{5ex} Top panel: Trajectories of two representative particles with the same initial conditions ($\alpha_{\rm i} \sim 0.075$) with $\kappa =1$ and $\kappa =\infty$ overplotted on a 2D slice of the 4-velocity component $\Gamma v_z$ of the flow. Bottom panel: Time evolution of particle energies, color coded with the instantaneous Lorentz factor probed, $\Gamma_{\rm pr}$ for the $\kappa =1$ particle. The particle gains a factor of $< 4$ in energy through stochastic acceleration and then goes through \emph{espresso} shots in the high-$\Gamma$ jet regions.}
  \includegraphics[width=0.5\textwidth]%
    {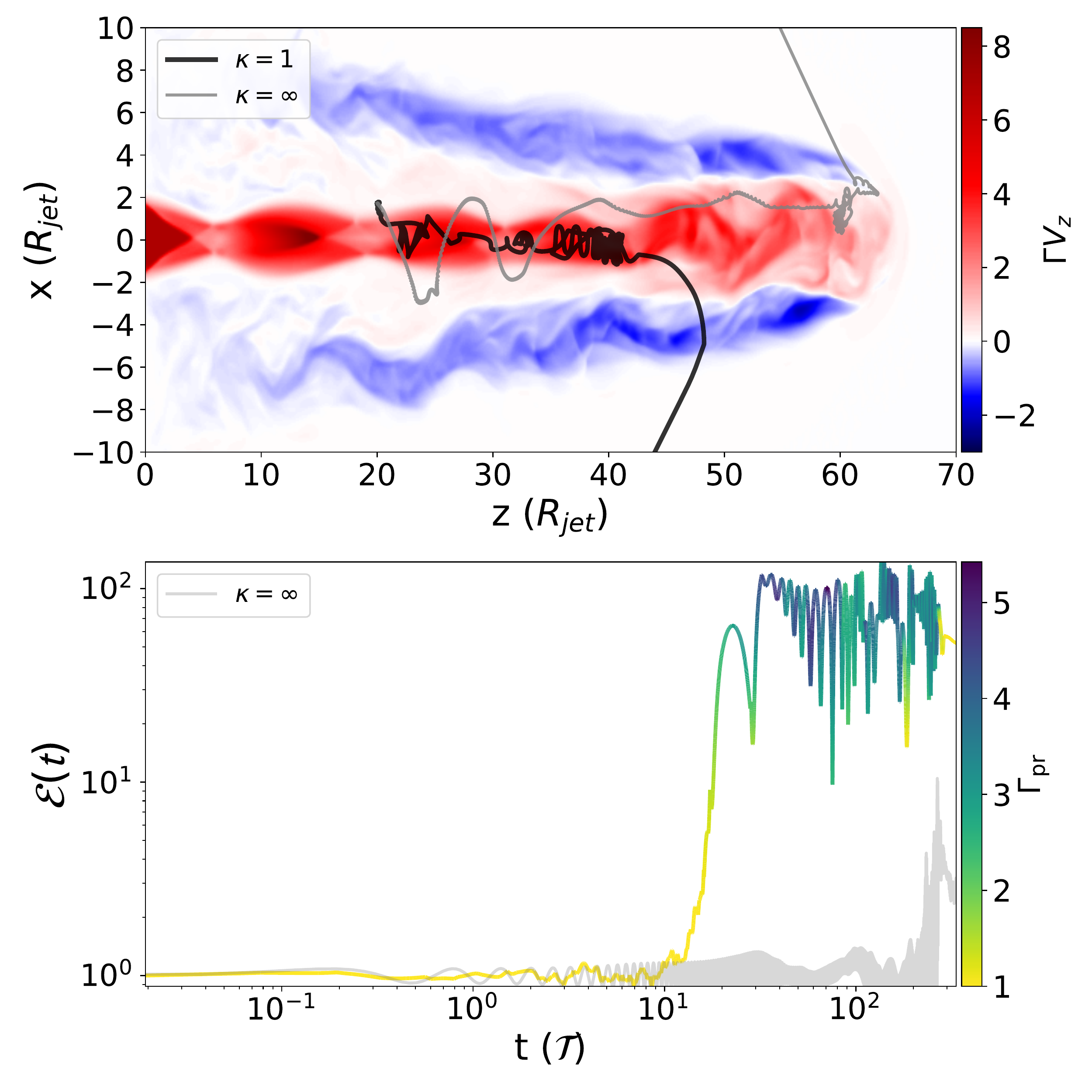}
    \label{traj-eg}
\end{SCfigure}

\subsection{SGS effect on particle trajectories and energy gain}
\begin{figure}
\centering
\includegraphics[width=1.0\textwidth,clip=false, trim= 0 0 0 0]{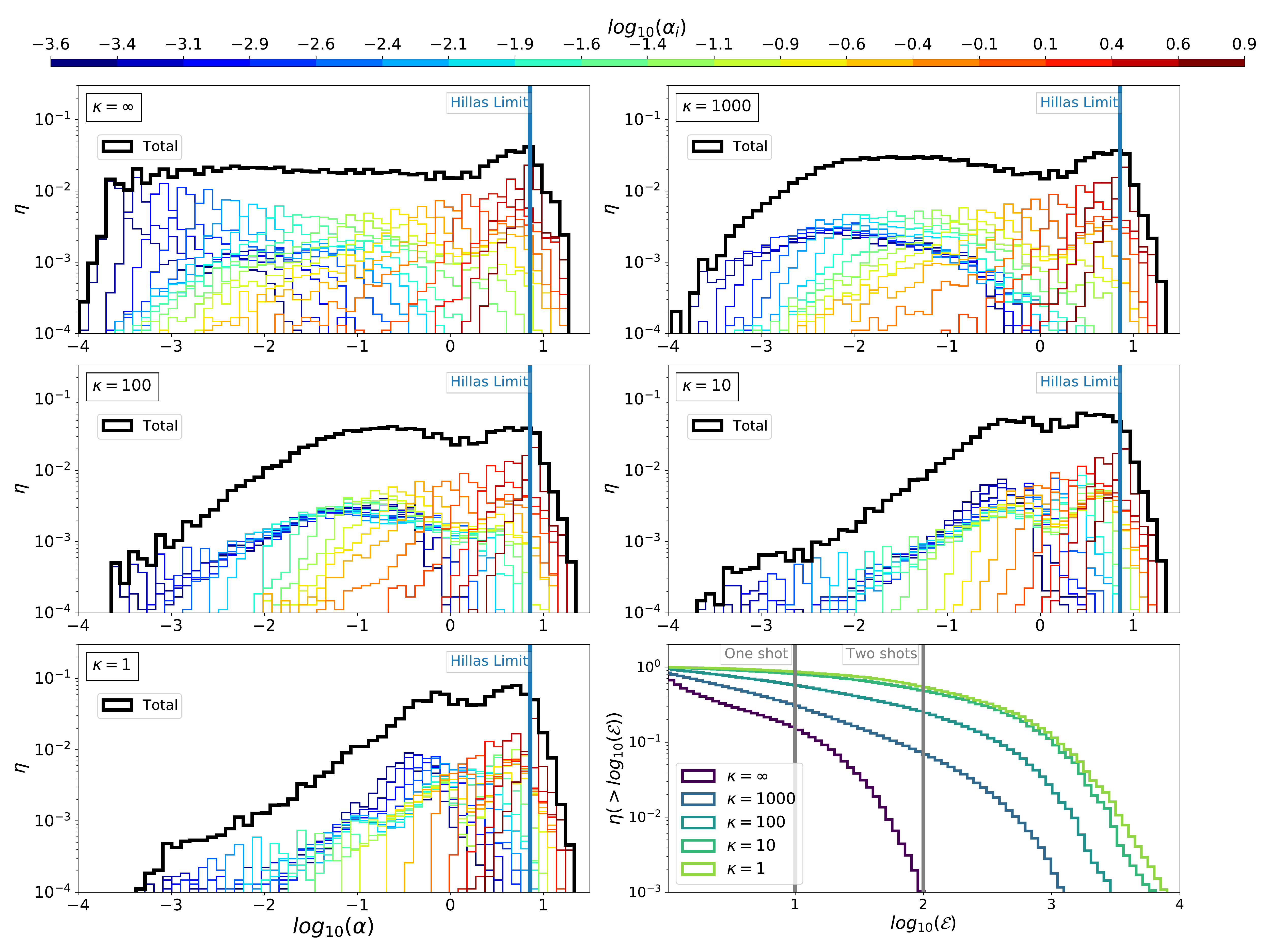}
\caption{Final spectrum of reaccelerated particles, assuming a seed spectrum $\propto \alpha_{\rm i}^{-1}$.
Colored histograms show the contribution of particles with initial gyroradii $\alpha_{\rm i}$, while the thick black line shows the cumulative spectrum.
Particles with $\alpha_{\rm i} \lesssim 1$ can undergo boosts $\lesssim 100\gg\geff^2$, while for $\alpha_{\rm i} \gtrsim 1$ the energy gain is smaller. The bottom right plot shows the cumulative distribution of the energy gains.}
\label{alpha}
\end{figure}

SGS is expected to enhance the diffusion of injected particles in and out of the highest-$\Gamma$ regions. 
In Figure~\ref{traj-eg}, we show a representative trajectory of a particle experiencing Bohm diffusion ($\kappa = 1$, black) along with a particle injected with the same initial conditions but with a scattering rate $\kappa = \infty$ (grey). 
The bottom panel shows the energy gain $\mathcal{E}$ of the particle with $\kappa = 1$ (color coded) as a function of its relativistic gyroperiod $\mathcal T \equiv \frac{2 \pi m \gamma_i}{q B_0}$ where $\gamma_i$ is the initial Lorentz factor, $q$ the charge, $B_0$ the reference value of the magnetic field, and $m$ the mass;
the grey curve shows the energy evolution of the particle with no SGS. 
The $\kappa = 1$ particle follows an irregular trajectory before entering the spine (where $\Gamma \gtrsim 2$) and experiencing \emph{espresso} gyrations, as opposed to the ordered gyrations we note in \citep{mbarek+19}. 
The $\kappa=\infty$ particle probes regions with $\Gamma < 2$ and exits the jet/cocoon system with $\mathcal E \sim 3$. 
We can then conclude that adding SGS enables stochastic acceleration, which can initially energize seeds, and help them to penetrate in the spine of the jet and get \emph{espresso} accelerated.

Then, we propagate $\sim 10^5$ test-particles in this jet with a variety of initial gyroradii $\mathcal R$ and positions, for the scattering rates discussed above. Particles are homogeneously and isotropically initialized in linearly spaced locations as discussed more in detail in \citep{mbarek+19}. 
Figure~\ref{alpha} shows the final spectra of accelerated particles produced with different SGS prescriptions. 
Here, $\alpha$ is defined as the particle gyroradius normalized to the jet radius such as $\alpha \left( E,q \right) \equiv \frac{\mathcal R \left( E,q \right) }{\rj}$ and we indicate with $\alpha_{\rm i}$  and $\alpha_{\rm f}$ the initial and final gyroradii;
the energy gain is then $\mathcal{E}\equiv\alpha_{\rm f}$/$\alpha_{\rm i}$.
The blue line delineates the Hillas criterion for the jet considered: it reads $\alpha\sim 8$ because the magnetic field equals $\sim 8 B_0$ in the spine region. 

It is important to stress that, regardless of the SGS rate, spectra are very similar at the highest energies, which implies that UHECRs are always \emph{espresso} accelerated. 
More importantly, SGS does not boost a larger fraction of the particles at the Hillas limit or to a larger maximum energy. 
However, low-energy seeds are considerably more likely to be boosted and energy gains of order of $\mathcal{E} \gtrsim 10^3$ are common for them. 
All these things considered, with SGS the final spectrum is significantly flatter than the injected one, by $\sim 0.9$ in slope for $\kappa = 1$, for instance.

\section{UHECR Photodisintegration and UHE Neutrinos}

\subsection{Particle Propagation}

We consider 5 different seed specie groups $s=$[H,He,C/N/O,Mg/Al/Si,Fe] with effective atomic number $Z_s=$[1, 2,7,13,26] and mass $A_s= $[1,4,14,27,56]. We parametrize their energy flux as $\phi_s(E) = K_s \left( \frac{E}{10^{12}~\rm eV}\right) ^{-q_s}$, where the normalizations are set based on the abundance ratios at $10^{12}$~eV such that $K_s $/$K_H \sim$ [1, 0.46, 0.30, 0.07, 0.14]. 
Note that, since propagation is rigidity dependent, a single integrated trajectory is representative of several different chemical species.

While propagating our particles, we consider proton-proton ($pp$), photomeson ($p\gamma$), and neutron decay from photodisintegration interactions as potential neutrino production mechanisms. 
In table~\ref{tab:interactions}, we show the interaction routes that lead to neutrino production and photodisintegration of heavy elements. 
The detailed explanations of particle interaction and attenuation losses will be presented in our upcoming paper [Mbarek et al.(2021)]. 
Finally, for studying the effects of photon fields on propagation, we set the SGS level to Bohm diffusion to maximize the number of injected particles within the spine of the jet and hence the probability of particle interactions.

Overall, at every time step we: (i) calculate the probability of interaction with the ambient medium and photon fields; 
(ii) keep track of each particle's atomic mass A and charge Z; and 
(iii) monitor secondary particle production and propagation.

\begin{table}
\begin{centering}
\resizebox{\textwidth}{!}{%
\begin{tabular}{ |c|c|c|c|c| } 
\hline
Particle & Process & Reactions & Neutrino Energy fraction \\
\hline
\hline
\multirow{2}{*}{Proton (p)} & photomeson ($p\gamma$) & $\ce{p}\text{+}\gamma \rightarrow \ce{n} \text{+}\pi^{\text{+}} \rightarrow \ce{n}\text{+}e^{\text{+}}\text{+}\nu_e\text{+}\nu_\mu\text{+} \bar{\nu}_\mu$ & p:$\nu \sim $ 20:1 \\
& proton-proton ($pp$) & $\ce{p}\text{+}\ce{p} \rightarrow \ce{p} \text{+} \ce{n} \text{+} \pi^{\text{+}} \rightarrow \ce{p} \text{+} \ce{n}\text{+}e^{\text{+}}\text{+}\nu_e\text{+}\nu_\mu\text{+} \bar{\nu}_\mu$ & p:$\nu \sim $ 20:1 \\
\hline
\multirow{2}{*}{Nucleus (N)} & photomeson ($p\gamma$) & $\ce{N}\text{+}\gamma \rightarrow \ce{n} \text{+}\pi^{\text{+}} \rightarrow \ce{n}\text{+}e^{\text{+}}\text{+}\nu_e\text{+}\nu_\mu \text{+} \bar{\nu}_\mu$ & N:$\nu \sim $ 20:1 \\ 
& photodisintegration & $\ce{^{A}N}\text{+}\gamma \rightarrow \ce{^{A-1}N}\text{+} \ce{n}\rightarrow \ce{^{A-1}N}\text{+} \ce{p}\text{+}e^-\text{+}\bar{\nu}_e$ & N:$\nu \sim 10^{3} $A:1 \\
\hline
\end{tabular}}
\caption{\label{tab:interactions} Neutrino production mechanisms along with their associated energy fractions. $A$ represents the atomic mass of the considered chemical specie.}
\end{centering}
\end{table}

\subsection{Photon Field Prescriptions}

On top of the MHD simulation, we prescribe external photon fields based mostly on the methods presented in \cite{murase+14}. 
We consider 5 main components for target photons: 
i) the nonthermal continuum from the blazar zone;
ii) the broad atomic line radiation (BLR) from cold gas clumps photoionized by UV and X-ray emission from the accretion disk;
iii) IR emission from the dusty torus;
iv) stellar light photons \cite{tanada+19}; and v) the cosmic microwave (CMB) background.
We consider here a photon field that maximizes particle interactions by choosing a very powerful blazar with bolometric luminosity $L_{\rm bol} = 10^{48}$erg/s \citep{murase+14}. 
Therefore, the results we obtain should be regarded as an \emph{upper limit} for neutrino production and photodisintegration. 
We finally note that the nonthermal contribution dominates particle interactions at all particle energies; 
this may have important repercussions for multimessenger astronomy since this component can be greatly enhanced, e.g., in AGN flares.

\begin{figure*}
\centering
\includegraphics[width=0.45\textwidth,clip=false,trim= 0 0 0 0]{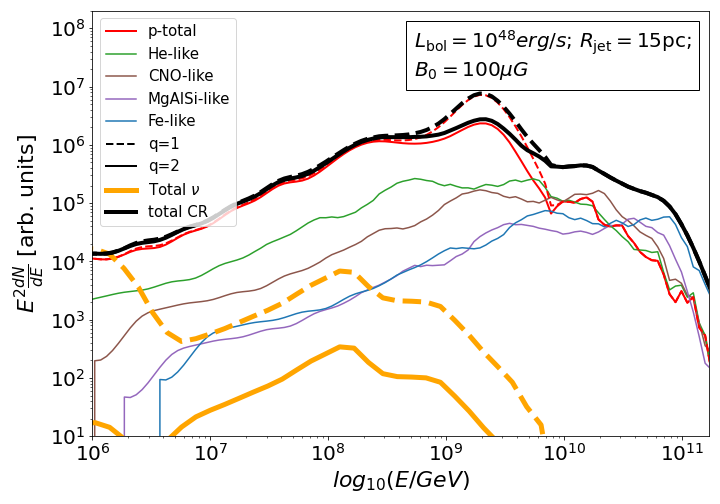}
\includegraphics[width=0.45\textwidth,clip=false,trim= 0 0 0 0]{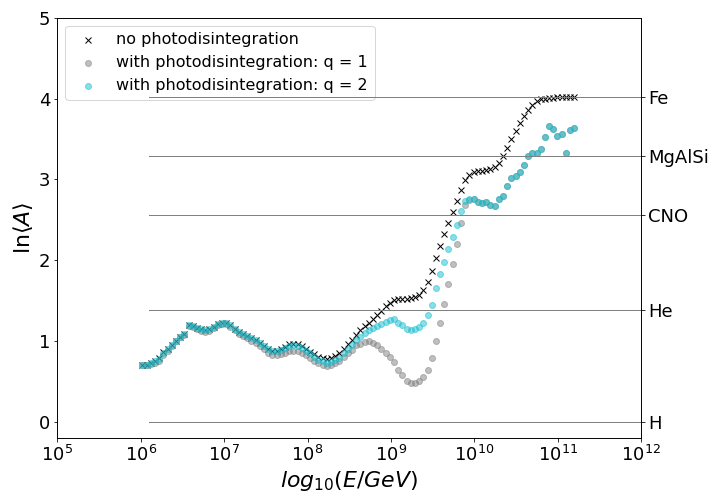}
\caption{Right Panel: Expected UHECR and UHE neutrino spectra assuming different injection spectral slopes $q$. 
He-like have atomic mass $ A\in$ [3,8]; CNO-like have $A \in$ [9,18]; MgAlSi-like have $A \in$ [19,35]; Fe-like have $ A\in$ [36,56]. 
Left Panel: Average atomic mass $A$ as a function of energy for different injection slopes. The atomic masses of the injected chemical species are marked with horizontal solid lines.}
\label{photo}
\end{figure*}

\subsection{Results}

\subsubsection{Effects of Losses on the UHECR Spectrum:}
In the left panel of Figure~\ref{photo}, we 
plot the UHECR spectrum in a blazar-like jet prescription along with contributions from the considered chemical species and neutrinos. 
Here, we note that the UHECR spectrum is not significantly affected by photodisintegration, which is further reiterated in the right panel of Figure~\ref{photo}, where we show the average atomic mass of the particles as a function of energy. 
The luminosity prescription we have set may magnify the effects of photodisintegration, however, the spectrum still gets heavier with increasing energy, which is consistent with Auger observations \citep{auger17_coll}. 
Additionally, we note the presence of a light-element bump---more prominent for flatter injection spectra---from reaccelerated secondary protons coincident with the EeV component, as also suggested by \cite{unger+15}. 
Overall, we note that \emph{espresso} acceleration is still consistent with UHECR energetics and chemical composition even in extreme blazar-like environments.

\subsubsection{Expected Spectrum of UHE Neutrinos from UHECR Interactions}

\begin{SCfigure}[][ht]
  \centering
  \caption{\protect\rule{0ex}{5ex} Expected source neutrino flux (black) from UHECR interactions in the \emph{espresso} framework assuming two different injection slopes $q=2$ (solid) and $q=1$ (dashed). We compare this flux to the expected cosmogenic neutrino flux (blue bands) \citep{batista+19}. UHECR data from Auger, KASCADE-Grande, and Telescope Array, as well as, neutrino data from IceCube are included for reference.}
  \includegraphics[width=0.45\textwidth]%
    {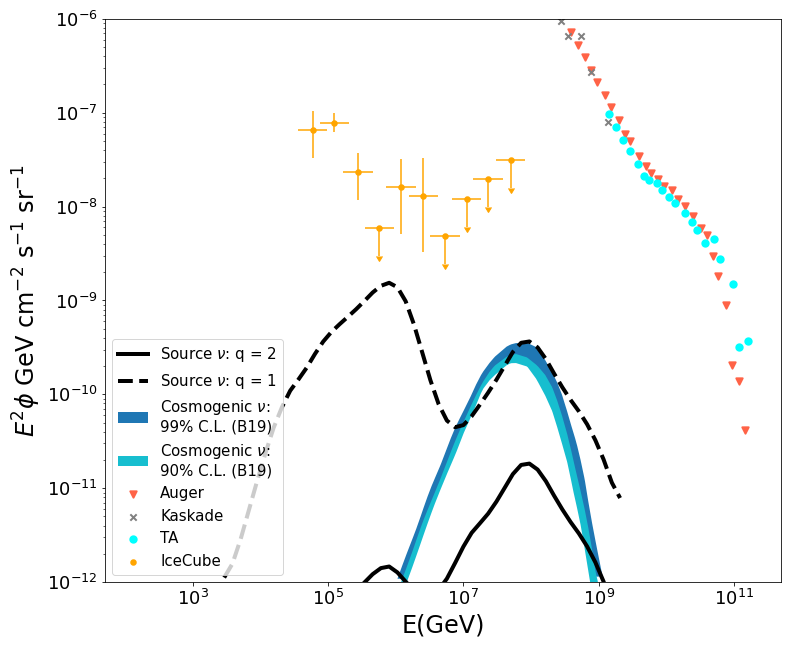}
    \label{data}
\end{SCfigure}

In Figure~\ref{data} we compare our expected UHE source neutrino emission with neutrino and UHECR data from IceCube, Auger, Telescope Array, and KASCADE-Grande based on a scaling of the relative normalizations of CR and neutrino spectra in the left panel of Figure~\ref{photo}. 
Namely, we take the relative UHECR to neutrino fluxes produced by our fiducial AGN source, and scale the expected neutrino flux with the observed UHECR one after accounting for the cosmological evolution of the sources. 
Our expected source UHE neutrino results are included in dotted and dashed black depending on the slope of seed spectrum, $\propto E^{-q}$. 
We find that neutrinos from neutron decay could contribute to IceCube's neutrino flux for rather flat injection spectra ($q=1$).
More importantly, these results suggest that the flux of source neutrinos at the EeV level may be comparable to that of cosmogenic neutrinos modeled based on an AGN source evolution with different confidence levels \citep{batista+19}.

\subsubsection{Neutrinos from Interactions of CRs with $E_{\rm CR} <10^{18}$~eV}

Below $\lesssim 10^{17}$ eV neutrinos can be produced in sources that are not necessarily UHECR sources, but we can still calculate the contribution of AGN jets to the flux measured by IceCube within the \emph{espresso} framework. 
Initially, for a given seed flux, we calculate the fraction of \emph{espresso}-reaccelerated particles and their corresponding energy gain. 
This is converted into a total luminosity in UHECRs for a fiducial AGN jet in a given galaxy;
then, from the fraction of energy carried by the produced neutrinos relative to the overall escaping particle energy, we compute the overall neutrino power. 
Finally, we then compare this expected luminosity with the expected neutrino power associated with FR-I and FR-II jets in the IceCube energy range and find that our calculated neutrino power is $\lesssim 10^{-4}-10^{-3}$ lower than what is expected by IceCube, even for a high-luminosity AGN and assuming Bohm diffusion for SGS.
Since these prescriptions maximize the expected neutrino yield, we can thereby conclude that AGN jets cannot be responsible for the the bulk of the neutrinos observed by IceCube, even if they were the main sources of UHECRs.

\section{Conclusions}

\begin{enumerate}
    \item Small-scale scattering, which we implenent as SGS in MHD simulations, may be important for the  acceleration of low-energy UHECRs. 
    Yet, the highest-energy CRs are invariably \emph{espresso}-accelerated and they reach the Hillas limit independently of the level of SGS.
    \item UHECRs accelerated in AGNs are not heavily affected by photodisintegration in their sources; since acceleration mainly occurs away from the jet base, even in powerful blazar we expect the spectrum of the highest-energy particles to get heavier with energy, as observed by Auger.
    \item The \emph{steady} neutrino emission from FR-Is/FR-IIs, the probable sources of UHECRs, cannot account for IceCube's astrophysical neutrinos.
    \item Source neutrinos with energies $> 10^{17}$eV could be associated with AGN jets, the UHECR acceleration sites, as their flux should be comparable to that of cosmogenic neutrinos.
    \item Nevertheless, since the jet nonthermal emission is the dominant photon background for CR interactions, neutrinos from AGN jets could correlate with $\gamma$-ray flares, which suggests that AGN flares are excellent multimessengers candidates.
\end{enumerate}

\bibliography{Total}
\end{document}